\documentclass[english,letterpaper,twocolumn,showpacs,prl,aps]{revtex4}
\usepackage{graphicx}
\usepackage{amssymb}

\makeatletter

\large  

\input epsf

\makeatother

\usepackage{babel}
\makeatother
\begin{document}

\title{Aharonov-Bohm effect in clean strong topological insulator wires}

\author{Eugene B. Kolomeisky$^{1}$ and Joseph P. Straley$^{2}$}

\affiliation
{$^{1}$Department of Physics, University of Virginia, P. O. Box 400714,
Charlottesville, Virginia 22904-4714, USA\\
$^{2}$Department of Physics and Astronomy, University of Kentucky,
Lexington, Kentucky 40506-0055, USA}

\begin{abstract}
Surface electrons of strong topological insulator wires acquire a Berry phase difference of $\pi$ on orbiting the wire.  This can be detected in response of clean wires (whose Fermi level is tuned to the Dirac point) to the presence of the Aharonov-Bohm flux.  Specifically, at half-odd integer applied flux (in units of $hc/e$), long wires undergo semimetal-semiconductor transitions characterized by logarithmically divergent susceptibility.  Associated with these are oscillations of magnetization (persistent current) that vanish both at integer and half-odd integer flux.  Additionally wires of arbitrary aspect ratio exhibit  conductance maxima at half-odd integer applied flux and minima at integer flux.  For long wires the maxima are sharp with their height approaching $e^{2}/h$.  Short wires are characterized by a universal conductivity  $e^{2}/\pi h$ attained in the disc limit.  
\end{abstract}


\maketitle

Three-dimensional strong topological insulators (TIs) are materials which are insulating in the bulk and have surface states with the graphene-like linear low energy-momentum dispersion relation that is characteristic of massless Dirac fermions \cite{TI_reviews}.  Therefore the physical properties of TI surfaces or wires resemble those of graphene \cite{RMP} or carbon nanotubes, respectively.  However, there are several important distinctions:

(i)  Graphene has two Dirac points per Brillouin zone while strong TIs have an odd number of surface Dirac points \cite{TI_reviews}.  Specifically, for the $(111)$ surface of $Bi_{2}Se_{3}$ \cite{BiSe} there is just one Dirac point.  Additionally, the surface electrons of TIs have their spin and momentum locked at right angle while the electrons in graphene are spin-degenerate.  Thus the electronic properties of the TI surface are more appropriately compared to a single valley of graphene without spin degeneracy \cite{TI_reviews}.  

(ii)  In graphene the physics of carbon atoms forming a honeycomb lattice automatically sets the Fermi energy at the Dirac point; this generally is not the case for surface states in TIs.  However, through a combination of surface and bulk chemical modification the surface Fermi level in TIs can be also tuned to the Dirac point \cite{tuning} which is what is assumed below.

Since TI wires can only conduct along the surface, they mimic hollow conducting cylinders, and then due to the Aharonov-Bohm (AB) effect \cite{AB}, their physical properties are sensitive to the presence of an axial magnetic field.  Due to the spin-momentum locking, the surface electrons acquire a Berry phase difference of $\pi$ \cite{disordered_wires} in addition to the AB phase while traveling around the wire.   This will have an experimental signature in various physical properties. The magneto-oscillations of conductance of TI wires have been observed \cite{experiment}; however, the Berry phase effect was not detected;  this was attributed to the effect of quenched disorder \cite{disordered_wires}.  In another experiment \cite{Hong} AB oscillations of the conductance consistent with the presence of the Berry phase were observed;  however, the authors stated that their wires were only ballistic around the circumference and not along the length of the wire.  More robust evidence of the Berry phase was reported only very recently \cite{recent} on intermediate length quasi-ballistic wires of $(Bi_{1.33}Sb_{0.67})Se_{3}$.      

In this paper we analyze the AB oscillations of conductance of ballistic TI wires of arbitrary aspect ratio and the AB magnetism of long wires.  Both of these contain fingerprints of the Berry phase which should help to identify the topological nature of TI surface states in future experiments.  Our treatment is modeled on previous analysis of AB effect in nanotubes \cite{KZS,KS}.        

Consider a TI wire of macroscopic length $L$ and circumference $W$ attached at its bases to conductive leads.  It is a fundamental property of the massless two-dimensional Dirac equation that the transmission probability of propagating modes in the leads is given by the formula \cite{Katsnelson}
\begin{equation}
\label{transmission}
T_{n}=\frac{1}{\cosh^{2}(q_{n}L)}
\end{equation}  
where the index $n$ labels the modes, and the azimuthal wave numbers $q_{n}$ are singled out by the behavior of the wave function when electron's position executes one full rotation about the wire axis.  At zero magnetic field one should choose $q_{n}=(2\pi/W)(n+1/2)$, $n=0, \pm 1, \pm 2, ...$ to include the electron's Berry phase of $\pi$.  The allowed azimuthal wave numbers $q_{n}$ will be modified by the presence of an axial magnetic field which is the heart of the AB effect.  Indeed, imagine there is an axial AB flux $\Phi$ entirely confined within the interior of the wire.  Adopting the gauge where the vector potential points in the azimuthal direction, we observe that the electrons experience a vector potential $A = \Phi/W$ whose effect can be accounted for via the Peierls substitution 
\begin{equation}
\label{wave_numbers}
q_{n}\rightarrow q_{n} - \frac{eA}{\hbar c} = \frac{2\pi}{W}\left (n + \frac{1}{2}-\phi\right ), n=0, \pm 1, \pm 2, ...
\end{equation}
where $\phi=\Phi/\Phi_{0}$ is the dimensionless flux measured in units of the flux quantum $\Phi_{0}= hc/e$. In present geometry the Peierls substitution is exact because in the natural coordinate system of the wire the vector potential has constant magnitude and direction.

The conductance of the TI wire (hereafter measured in units of the half conductance quantum $e^{2}/h$) is given by the Landauer  transmission formula \cite{Landauer}
\begin{eqnarray}
\label{TI_conductance_formula}
G(\phi)&=& \sum_{n=-\infty}^{\infty}T_{n}\nonumber\\
&=&\sum_{n=-\infty}^{\infty}\frac{1}{\cosh^{2}[(n+1/2-\phi)2\pi L/W]}
\end{eqnarray}
We see that the conductance is an even periodic function of dimensionless flux $\phi$ with unit period having maxima at half-odd integer flux and minima at integer flux.  The character of this dependence is exponentially sensitive to the wire aspect ratio $W/L$.      

In the disc limit ($W/L\rightarrow\infty$), the sum (\ref{TI_conductance_formula}) can be approximated by an integral with the result $G(\phi)\approx(1/\pi)(W/L)$ thus leading to a finite universal conductivity $\sigma_{0}=GL/W=1/\pi$ which is exactly $1/4$ of conductivity of clean graphene sheet \cite{Katsnelson}.   The fingerprints of both the Berry and AB phases are lost because finite-size quantization of the azimuthal motion is irrelevant in the disc limit.  The conclusion of universal conductivity of $1/\pi$ also applies to transport along the short direction of a TI ribbon of length $L$ and width $W \gg L$ with arbitrary boundary conditions at the ribbon's edges.    

In the limit of a wire of infinite length ($W/L\rightarrow0$), the discreteness of the azimuthal wave numbers is relevant, and the wire is insulating ($G(\phi)=0$) unless the applied flux is exactly a half-odd integer. In the latter case the wire is conductive with unit ($e^{2}/h$) conductance.  This result, anticipated in previous work \cite{disordered_wires}, is due to the presence of a single conductive mode $q_{n}=0$, $n+1/2-\phi=0$ which for an infinitely long wire prevents a single term of the mode sum (\ref{TI_conductance_formula}) from vanishing.  We thus conclude that half-odd integer fluxes represent loci of the semiconductor-semimetal phase transitions.   Analogous transitions are known to exist in carbon nanotubes \cite{Datta, KZS}.          

For finite aspect ratio $W/L\ll 1$ the semiconductor-semimetal transitions are no longer sharp.  If the difference between applied flux and nearest half-odd integer flux $\bigtriangleup \phi$ satisfies the condition $(2\pi L/W) |\bigtriangleup \phi| \lesssim 1$, the conductance will be close to its ideal value $G=1$;  otherwise it will be exponentially suppressed.  The width of the conductance peak is thus of the order $W/2\pi L$.  The practical consequence is that conductance peaks of approximately $e^{2}/h$ amplitude must be already visible for relatively short wires with $L\simeq W$ which agrees with experimental observations \cite{recent}.      

Since the conductivity of long wires at half-odd integer flux is divergent ($\sigma=GL/W\simeq L/W$) while the conductivity in the disc limit is universal ($\sigma_{0}=1/\pi$) one would expect that the conductivity at half-odd integer flux is a monotonically decreasing function of the aspect ratio $W/L$.   Having the applied flux deviate from a half-odd integer value by $\bigtriangleup \phi$ qualitatively changes the aspect ratio dependence of conductivity because long wires are now insulating while short wires remain conducting with universal conductivity of $1/\pi$.  The conductivity of a long wire $\sigma \simeq 4(L/W)\exp(-4\pi L|\bigtriangleup \phi||/W)$  vanishes in the $W/L\rightarrow 0$ limit but has a maximum for $W/L\simeq 4\pi |\bigtriangleup \phi|$.  As one approaches a half-integer flux $\bigtriangleup \phi \rightarrow 0$, the conductivity maximum shifts towards $W/L=0$ and grows in amplitude.  We conclude that if the applied flux is sufficiently close to a half-odd integer flux the conductivity is a non-monotonic function of the aspect ratio $W/L$ with a maximum whose position is set by the vicinity to the case of a half-odd integer flux.   As shown in Figure 1, when the applied flux is sufficiently far away from half-odd integer flux, there is no maximum at all and the conductivity is a monotonically increasing function of the aspect ratio $W/L$.  
\begin{figure}
\includegraphics[width=1.0\columnwidth, keepaspectratio]{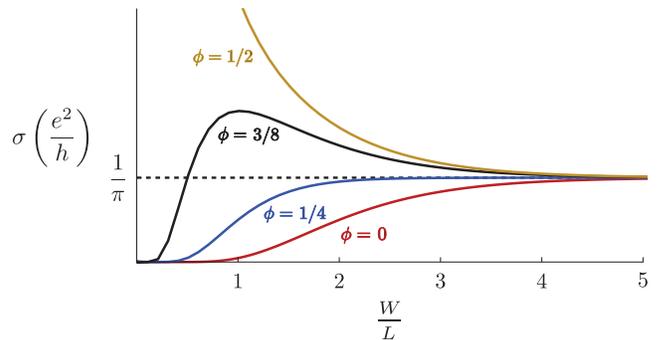} 
\caption{(Color online) Conductivity of strong TI wires versus their aspect ratio $W/L$ for different values of the applied AB flux $\phi$ (chosen to be in the $0 \leqslant \phi \leqslant 1/2$ range).  The boundary between monotonic and non-monotonic aspect ratio dependence of the conductivity is at $\phi_{c}=1/4$.}
\end{figure}

When $W/L$ is large it is useful to use the Poisson summation formula \cite{LL5} to sum over $n$ in Eq.(\ref{TI_conductance_formula}) with the results for the conductance
\begin{equation}
\label{Poisson_chirality_conductance}
G(\phi)=\frac{W^{2}}{2L^{2}}\sum_{n=-\infty}^{\infty}\frac{n\cos[2\pi n(\phi-1/2)]}{\sinh(\pi nW/2L)}
\end{equation}
and conductivity
\begin{equation}
\label{Poisson_chirality_conductivity}
\sigma(\phi)=\frac{1}{\pi}+ \frac{W}{L}\sum_{n=1}^{\infty}\frac{n\cos[2\pi n(\phi-1/2)]}{\sinh(\pi nW/2L)}
\end{equation}
respectively.  We now observe that for arbitrary aspect ratio $W/L$ the effect of the AB flux manifests itself in periodic oscillations of the conductivity about $\sigma_{0}=1/\pi$;  the Berry phase sets the phase of the oscillation while the aspect ratio determines the amplitude of the conductivity oscillation.  

In the disc limit, $W/L\rightarrow\infty$, the sum in Eq.(\ref{Poisson_chirality_conductivity}) is well-approximated by the constant and $n=1$ terms with the result
\begin{equation}
\label{Poisson_flux_conductivity_ring}
\sigma(\phi)\approx\frac{1}{\pi}- \frac{2W}{L}\exp \left (-\frac{\pi W}{2L}\right )\cos(2\pi \phi)
\end{equation}
We observe that as $W/L\rightarrow\infty$ the limiting conductivity of $1/\pi$ is approached from below if $0\leqslant \phi \leqslant1/4$ and from above if $1/4<\phi \leqslant1/2$.  This demonstrates that $\phi_{c}=1/4$ as suggested by Figure 1.  

Our results summarizing the effects of the AB flux and Berry phase on the conductance of TI wires of various aspect ratios are shown in Figure 2.  Wires studied in recent experiment \cite{recent} had the aspect ratio $W/L\simeq 1$;  reported flux dependence of the conductance is qualitatively similar to our result shown in Figure 2.
\begin{figure}
\includegraphics[width=1.0\columnwidth, keepaspectratio]{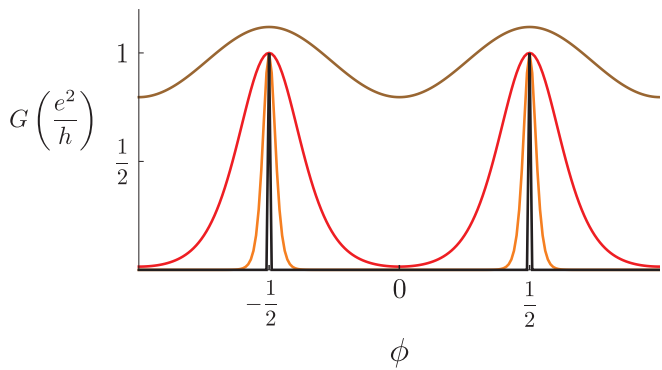} 
\caption{(Color online) Conductance of strong TI wires versus applied AB flux $\phi$ for different values of their aspect ratio $W/L$.  The latter are color-coded according to $W/L=0.0018$ (black), $W/L=1/5$ (orange), $W/L=1$ (red), and $W/L=3$ (brown).}
\end{figure}   

The presence of the Berry phase also manifests itself in the magnetic response of TI wires to the presence of AB flux.  In computing this response we limit ourselves to the case of long $L \gg W$ wires where the effect is strongest.  The surface band structure of the TI wire in the presence of the AB flux is given by
\begin{equation}
\label{band_structure}
E_{n}(q_{z})=\pm \hbar v_{F}\left [q_{z}^{2}+\left (\frac{2\pi}{W}\right )^{2}(n+\frac{1}{2}-\phi)^{2}\right ]^{\frac{1}{2}}
\end{equation}
where the upper and lower signs refer to the empty and filled bands, $v_{F}$ is the Fermi velocity, $q_{z}$ is the axial projection of the wave vector, and the second term inside the brackets is the square of the azimuthal projection of the wave vector (\ref{wave_numbers}).  The immediate consequence of Eq.(\ref{band_structure}) is that the energy gap at $q_{z}=0$ can be periodically closed by half-odd integer AB flux $\phi$ thus triggering the already mentioned semiconductor-semimetal transitions.

The occupied negative energy states of the spectrum (\ref{band_structure}) contribute into the ground-state energy the quantity
\begin{eqnarray}
\label{E_K}
\mathcal{E}(\phi)&=&-\hbar v_{F} \sum_{n=-\infty}^{\infty}\int_{-\infty}^{\infty}\frac{Ldq_{z}}{2\pi}\nonumber\\
&\times& \left [q_{z}^{2}+\left (\frac{2\pi}{W}\right )^{2}(n+\frac{1}{2}-\phi)^{2}\right ]^{\frac{1}{2}}
\end{eqnarray}
The magnetic moment $\mathcal{M}(\phi)$ and differential susceptibility $\chi(\phi)$ are determined by differentiation of Eq.(\ref{E_K}):
\begin{equation}
\label{M_K_chi_K}
\mathcal{M}(\phi)=-\frac{W^{2}}{4\pi\Phi_{0}} \frac{\partial \mathcal{E}}{\partial \phi}=\frac{W^{2}}{4\pi c}\mathcal{I},~~~~~~~\chi(\phi)=\frac{W^{2}}{4\pi\Phi_{0}}\frac{\partial \mathcal{M}}{\partial \phi}
\end{equation}
where $\mathcal{I}=-(c/\Phi_{0})\partial \mathcal{E}/\partial \phi$ is the persistent current.

The apparent divergence of Eq.(\ref{E_K}) is fictitious because the  expression for the spectrum (\ref{band_structure}) is only applicable at low energy, and the sum and integral should only be over wavevectors within the first Brillouin zone.   Analysis of a nearly identical problem \cite{KS} shows that if we are only interested in the flux-dependent part of the ground-state energy $\mathcal{E}^{'}(\phi)$ determining measurable quantities (\ref{M_K_chi_K}), then Eq.(\ref{E_K}) is sufficient, with a universal result for the ground-state energy:
\begin{equation}
\label{AB_energy}
\mathcal{E}^{'}(\phi)=\frac{\hbar v_{F} L}{\pi W^{2}}\sum_{n=1}^{\infty}\frac{\cos[ 2\pi n(\phi-1/2)]}{n^{3}}
\end{equation}
We see that the AB flux controls both the magnitude and sign of the result; the energy has minima at $\phi$ integer and maxima at $\phi$ half-odd integer.  The magnetic moment and susceptibility of the wire follow from Eqs.(\ref{M_K_chi_K}) as
\begin{equation}
\label{M_K}
\mathcal{M}(\phi)=\frac{\hbar v_{F} L}{2\pi \Phi_{0}}\sum_{n=1}^{\infty}\frac{\sin [2\pi n(\phi-1/2)]}{n^{2}}
\end{equation}
\begin{equation}
\label{chi_K}
\chi(\phi)= -\frac{\hbar v_{F} W^{2}L}{4\pi \Phi_{0}^{2}}\ln[2|\cos (\pi \phi)|]
\end{equation}
which are also shown in Figure 3.  Both quantities are proportional to the wire length $L$; the magnetic moment is independent of the circumference $W$, while the susceptibility, proportional to the wire volume, is logarithmically divergent at $\phi$ half-odd integer.  This divergence translates into a weak non-analyticity at half-odd integer values of $\phi$ for the energy (\ref{AB_energy}) and magnetic moment (\ref{M_K}), with the latter vanishing both at integer and half-odd integer $\phi$.    The AB response of the wire is diamagnetic for $0\leqslant \phi < 1/6$ and paramagnetic for $1/6 < \phi \leqslant 1/2$. Apart from the overall shift by a half flux quantum, Eqs.(\ref{M_K}) and (\ref{chi_K}) describes the AB magnetism per spin per valley of a semimetallic carbon nanotube \cite{Ando}. 

\begin{figure}
\includegraphics[width=1.0\columnwidth, keepaspectratio]{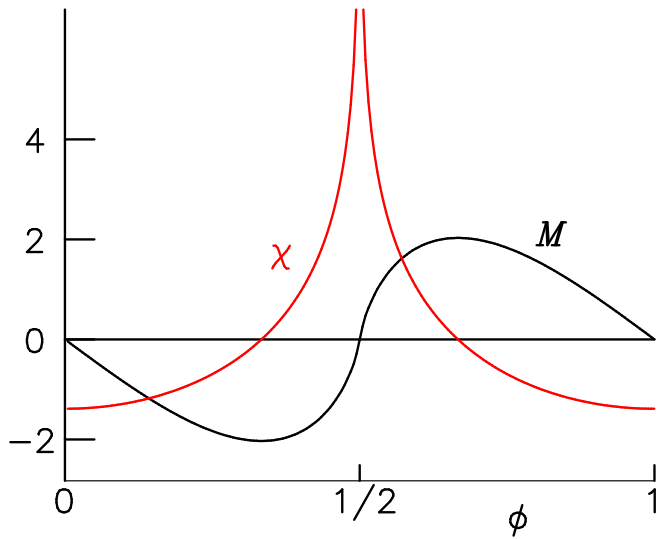} 
\caption{(Color online)   Magnetization (in units of $\hbar v_{F}L/2\pi \Phi_{0}$, black) and susceptibility (in units of $\hbar v_{F}W^{2}L/4\pi\Phi_{0}^{2}$, red) of a long TI wire, given by Eqs.(\ref{M_K}) and (\ref{chi_K}) as functions of the applied AB flux $\phi$.}
\end{figure}
Experimental detection of the Berry phase effect in TI wires is easier than study of the semiconductor-semimetal transitions in a nanotube.   Available nanotubes have very small diameter, so that enclosing one flux quantum requires a very large magnetic field and makes the observation of one full AB oscillation impossible.  In practice this limits the possibility of observation of the semiconductor-semimetal transition to semiconductor tubes \cite{KZS}.  Such a limitation does not exist for TI wires (as was recently demonstrated \cite{recent}), because they can easily be made wider.  Moreover sufficiently short nanotubes are not experimentally available which makes it challenging to detect the regime of universal conductivity.  On the other hand, the aspect ratio of TI wires can be made arbitrary thus making it possible to test the entire dependence of transport on wire's aspect ratio.  Due to the larger size of the samples, experimental detection of the magnetic properties of TI wires also seems to be feasible.

This work was supported by the US AFOSR grant FA9550-11-1-0297.

\end{document}